# Spin and orbital ordering in TlMnO$_3$: Neutron diffraction study

Dmitry D. Khalyavin,[1,*] Pascal Manuel,[1] Wei Yi,[2,3] and Alexei A. Belik[3,4]

[1]*ISIS Facility, Rutherford Appleton Laboratory, Chilton, Didcot, Oxon, OX11 0QX, United Kingdom*
[2]*Institute of Physics and Beijing National Laboratory for Condensed Matter Physics, Chinese Academy of Sciences, Beijing 100190, China*
[3]*International Center for Materials Nanoarchitectonics (WPI-MANA), National Institute for Materials Science (NIMS), Namiki 1-1, Tsukuba, Ibaraki 305-0044, Japan*
[4]*Research Center for Functional Materials, National Institute for Materials Science (NIMS), Namiki 1-1, Tsukuba, Ibaraki 305-0044, Japan*

Crystal and magnetic structures of the high-pressure stabilized perovskite phase of TlMnO$_3$ have been studied by neutron powder diffraction. The crystal structure involves two types of primary structural distortions: $a^+b^-b^-$ octahedral tilting and antiferrodistortive type of orbital ordering, whose common action reduces the symmetry down to triclinic $P\bar{1}$. The orbital pattern and the way it is combined with the octahedral tilting are different from the family of LnMnO$_3$ (Ln = lanthanide or Y) manganites who share with TlMnO$_3$ the same tilting scheme. The experimentally determined magnetic structure with the $k = (1/2, 0, 1/2)$ propagation vector and $P_S\bar{1}$ symmetry implies anisotropic exchange interactions with a ferromagnetic coupling within the $(1, 0, \bar{1})$ planes and an antiferromagnetic one between them (A type). The spins in the primary magnetic mode were found to be confined close to the $(1, 0, \bar{1})$ plane, which underlines the predominant role of the single ion anisotropy with the local easy axes of Mn$^{3+}$ following the Jahn-Teller distortions of the octahedra. In spite of the same octahedral tilting scheme in the perovskite structures of both LnMnO$_3$ and TlMnO$_3$ manganites, a coupling of the secondary ferromagnetic component to the primary A-type spin configuration through antisymmetric exchange interaction is allowed in the former and forbidden in the latter cases.

## I. INTRODUCTION

Complex manganese oxides with perovskite structures exhibit a variety of fascinating physical properties which have been the focus of intensive studies for the last several decades. Among these properties, colossal magnetoresistance [1–3] and spin-driven ferroelectricity [4–6] are particularly useful, making these materials to be very topical in the field of condensed matter physics and materials science. The prototype members of the family are LnMnO$_3$ manganites which contain Jahn-Teller active Mn$^{3+}$ cations in octahedral coordination. The orbital degree of freedom is the crucial ingredient of their rich physical properties [7], and cooperative Jahn-Teller distortions take place in all these compounds at very high temperatures, typically above 750 K [8]. Magnetic ordering establishes at substantially lower temperatures, usually below 140 K [9], and is strongly influenced by the ordering of the occupied/empty electronic states (orbital ordering) controlled by the Jahn-Teller distortions [10,11]. In the manganites with relatively large Ln=La-Gd, the spin arrangement is dominated by the nearest-neighbor Heisenberg exchange, resulting in a so-called A-type antiferromagnetic structure [9]. This type of magnetic ordering implies anisotropic exchange interactions with ferromagnetic in-plane and antiferromagnetic out-of-plane coupling imposed by the orbital ordering. In the perovskites with Ln of smaller size, the spin ordering is affected by the next nearest neighbor exchange, stabilizing incommensurate [12–14] or E-type magnetic structures [9,15], and the role of the orbital degree of freedom is less pronounced.

Another example of a simple perovskite with manganese being in the 3+ charge state is BiMnO$_3$, which exhibits a ferromagnetic insulating behavior [16]. This ground state is believed to be stabilized by a peculiar orbital pattern with competing ferromagnetic and antiferromagnetic exchange [17]. Both LnMnO$_3$ and BiMnO$_3$ manganites under a hole doping with alkaline-earth metals exhibit a delicate balance between electronic, spin, and lattice degrees of freedom, resulting in a variety of exotic charge/orbital/spin ordered phases [1–3]. In this respect, the most studied case is the CE phase in the half-doped manganites [18,19]. The phase combines all three instabilities (charge, orbital, and spin) and can be melted by an external magnetic field, giving rise to the effect of colossal magnetoresistance [2,3,7].

Very recently a new member of the Mn$^{3+}$-based perovskite TlMnO$_3$ has been reported with a triclinic crystal structure and a clear signature of Jahn-Teller distortions [20]. The material has a great potential to become a new playground in the manganites family and exhibits a long-range magnetic order below $T_N \sim 92$ K. The magnetic properties have been studied with specific heat, magnetization, and Mossbauer spectroscopy methods combined with state of the art density functional calculations. The orbital degree of freedom has been suggested to play a crucial role in the formation of the magnetic ground state of TlMnO$_3$, and an original A-type spin order, distinct from that in LnMnO$_3$, has been predicted based on the first principle calculations. In the present work, we experimentally studied the magnetic structure of TlMnO$_3$ using neutron diffraction technique. Our study confirmed the presence of anisotropic exchange interactions in the system, predicted theoretically, and revealed that the Jahn Teller distortions are a key factor in establishing the magnetocrystalline anisotropy. Using parameters obtained in the Rietveld refinements of the crystal and magnetic structures, we decomposed the distortions into a set of symmetry-adapted displacement and magnetic modes. This node decomposition allowed us to identify the primary and secondary order parameters and better understand

---

*dmitry.khalyavin@stfc.ac.uk

the role of the structural distortions in the magnetic properties of manganites.

## II. EXPERIMENTAL SECTION

The polycrystalline samples of TlMnO$_3$ were prepared under high pressure (6 GPa) and high temperature (1500 K) conditions as described in Ref. [20]. The neutron powder diffraction data were collected at the ISIS pulsed neutron and muon facility of the Rutherford Appleton Laboratory (UK), on the WISH diffractometer located at the second target station [21]. The sample (∼100 mg) was loaded into a cylindrical 3-mm-diameter vanadium can and measured in the temperature range of 1.5–120 K (step 5 K, exposition time 20 minutes) using an Oxford Instrument Cryostat. Rietveld refinements of the crystal and magnetic structures were performed using the Fullprof program [22] against the data measured in detector banks at average $2\theta$ values of 58°, 90°, 122°, and 154°, each covering 32° of the scattering plane. Group-theoretical calculations were done using ISOTROPY [23], ISODISTORT [24], and Bilbao Crystallographic Server (AMPLIMODES: Symmetry mode analysis [25] and Magnetic Symmetry and Applications [26]) software.

## III. RESULTS AND DISCUSSION

### A. Crystal structure

The nuclear structure of TlMnO$_3$ has been refined in the triclinic $P\bar{1}$ space group using the atomic coordinates reported by Yi *et al.* [20] as the starting model. The model provides a very good fitting quality (Fig. 1) for the data sets collected over the whole temperature range of 1.5 K ⩽ T ⩽ 120 K. The unit cell parameters, Mn-O bond lengths and Mn-O-Mn bond angles as a function of temperature are shown in Fig. 1S– 3S of the Supplemental Material [27]. The unit cell parameters demonstrate a small kink near $T_N \sim 92$ K, where a magnetic phase transition (discussed in the next section) takes place, indicating a lattice relaxation due to magnetoelastic coupling. The Mn-O distances are weakly sensitive to the magnetic transition, whereas the bond angles tend to increase below $T_N$. The superexchange via oxygen is known to be very sensitive to electron hopping integrals controlled by the bond angles [28] and the observed behavior, therefore, is a natural response of the lattice to optimize the magnetic interactions in the system.

The structural parameters refined at 1.5 K are summarized in Table 1S of the Supplemental Material [27]. They were used to decompose the structure into a set of symmetry adapted displacement modes (mode decomposition) specified in terms of irreducible representations (irreps) of the parent $Pm\bar{3}m$ space group [24,25]. This procedure is a natural way to classify and quantify the distortions involved and allows differentiating between primary and secondary distortive modes based on their symmetries and amplitudes [24,25,29]. The result of the decomposition, with a short list of the displacement modes essential for our further discussion, is summarized in Table I. A full list of the modes involved can be found in Table 2S of the Supplemental Material [27].

The displacement modes with the largest amplitudes belong to the $R_4^+$ and $M_3^+$ irreps. These distortions involve out-of-phase and in-phase octahedral tilting, respectively, and are common for many perovskite systems [30,31]. The order parameter transformed by the $R_4^+$ irrep is allowed to take a general direction $(r_1, r_2, r_3)$ in the triclinic structure, but the absolute values of the amplitudes associated with the first and the second components are practically equal to each other and substantially bigger than the third one ($|r_1| \approx |r_2| \gg |r_3|$). This implies that with a good approximation, the tilting pattern is $a^-b^+a^-$ specified in terms of the Glazer's notation [30] (in the following discussion we will use a more common $a^+b^-b^-$ reference to this tilting pattern). In the setting adopted for the triclinic cell, the out-of-phase and in-phase octahedral tilting takes place about the $a$ and $b$ axes, respectively. A combination of the out-of-phase and in-phase tilting results in a coupling of displacement modes with the symmetry of $X_5^+(x_1, x_2, 0, 0, 0, 0)$ order parameter through the trilinear coupling term:

$$x_1 r_2 m_2 - x_2 r_1 m_2 + x_3 r_1 m_3 - x_4 r_3 m_3 + x_5 r_3 m_1 - x_6 r_2 m_1 \quad (1)$$

which is reduced down to $x_1 r_2 m - x_2 r_1 m$ for the $(0, m_2 = m, 0)$ direction in the $M_3^+$ representation space (Table I). Thus, the distortions associated with the $X_5^+$ irrep are secondary order parameters which are expected to be present in any perovskite structure involving both in-phase and out-of-phase octahedral tilting. The substantial amplitudes of these displacement modes in TlMnO$_3$, however, indicate that they play an essential role in the stability of the structure. As it will be discussed below, the $X_5^+$ distortion modes are also important for magnetic properties and are responsible for the weak ferromagnetism in LaMnO$_3$.

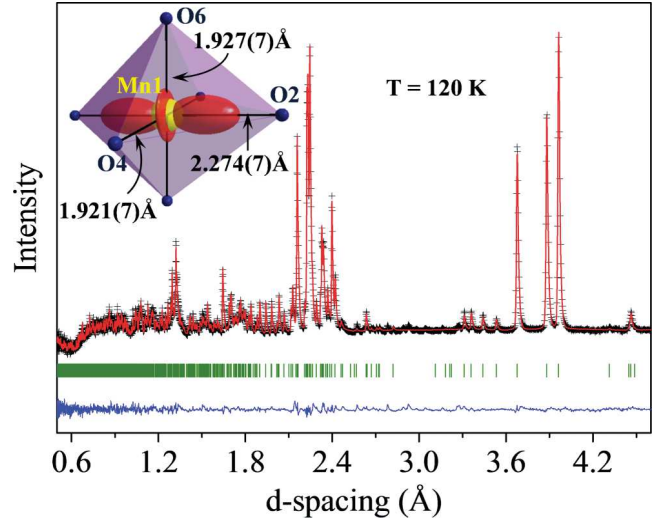

FIG. 1. A typical Rietveld refinement ($R_{Bragg} = 4.13\%$) of the high-resolution backscattering neutron diffraction pattern. The cross symbols (black) and solid line (red) represent the experimental and calculated intensities, respectively, and the line below (blue) is the difference between them. Tick marks indicate the positions of Bragg peaks in the $P\bar{1}$ space group. Inset shows the Mn-O bond distances and $d_{z^2}$ orbital, stabilized by the local Jahn-Teller distortion, for an arbitrary chosen octahedron, coordinating the Mn1 crystallographic site (the bond lengths for other manganese octahedra and their temperature variation can be found in Fig. 2S of the Supplemental Material [27]).

The $a^+b^-b^-$ tilting pattern reduces the symmetry of perovskite structure down to orthorhombic $Pnma$, and it is the common tilting scheme for the family of manganites LnMnO$_3$ [9]. We used the atomic coordinates for LaMnO$_3$, taken from the MAGNDATA database (#01) [26], to make the mode decomposition of the orthorhombic structure (see Table I for the short list and Table 5S of the Supplemental Material [27] for the full list of the modes involved). The Mn cations take the 4$b$ crystallographic site in the $Pnma$ space group with the site symmetry $\bar{1}$, which removes the degeneracy between $d_{x^2-y^2}$ and $d_{z^2}$ orbitals of Mn$^{3+}$. Thus, the $Pnma$ structure allows an orbital ordering without a further symmetry breaking, and strong Jahn-Teller distortions in LnMnO$_3$ take place as isostructural phase transitions [8]. The orbital pattern compatible with the $Pnma$ symmetry is shown in Fig. 2 (left). It implies that the distortion of octahedra due to stabilization of $d_{z^2}$ orbital (Jahn-Teller distortion) takes place within the (0,1,0) planes which are stacked in the same way along the $b$ axis (ferrotype stacking). This orbital ordering has the symmetry of the $M_2^+(0,a,0)$ order parameter (Table I) and its coupling to the octahedral tilting has been discussed by Carpenter and Howard [32]. In the triclinic structure of TlMnO$_3$, the $M_2^+$ distortions are negligibly small (Table I). The MnO$_6$ octahedra, however, are strongly irregular, pointing to the presence of Jahn-Teller distortions (see inset of Fig. 1 and Fig. 2S of the Supplemental Material [27]). These distortions are associated with the $R_3^+(0,b)$ mode which has a substantial amplitude practically identical to the absolute value of the $M_2^+$ Jahn-Teller mode in LaMnO$_3$ (Table I). The $R_3^+(0,b)$ displacement mode, whose Jahn-Teller nature has been pointed out in Ref. [32], implies a three-dimensional antiferrodistortive orbital ordering (G-type) shown in Fig. 2 (right). The orbital pattern involves an antiferrotype of stacking of the $(1,0,\bar{1})$ planes with the Jahn-Teller distortions of octahedra.

TABLE I. Decomposition of the $P\bar{1}$ structure of TlMnO$_3$ (unit cell parameters: $a_t = 5.41987(9)$ Å, $b_t = 7.9250(1)$ Å, $c_t = 5.27683(9)$ Å, $\alpha = 87.933(1)°$, $\beta = 86.804(1)°$, and $\gamma = 89.338(1)°$) and $Pnma$ structure of LaMnO$_3$ (unit cell parameters: $a_o = 5.7461(2)$ Å, $b_o = 7.6637(4)$ Å, and $c_o = 5.5333(2)$ Å) in respect of the symmetrized displacive modes of the parent cubic $Pm\bar{3}m$ perovskite structure. The column "Irrep ($\mathbf{k}$)" shows the irreducible representations of the $Pm\bar{3}m$ space group and the arms of the wave vector star involved (in brackets). The column "Order parameter" lists the order parameter directions in the corresponding irreducible subspaces (the same symbol in different positions indicates equal order parameter components). The column "Site irrep(var)" displays the point-group symmetry irreps of the local Wyckoff positions and one of the free variables from the order parameter directions (in brackets). "Amplitude" shows the amplitudes of the displacive modes (in Å). We provide the standard supercell-normalized amplitudes ($A_S$) defined as the square root of the sum of the squares of the mode induced atomic displacements within the primitive supercell.

| Irrep ($\mathbf{k}$) | Order parameter | Site irrep(var) | Amplitude |
|---|---|---|---|
| | $P\bar{1}$ structure of TlMnO$_3$ | | |
| | Tl displacement | | |
| $X_5^+(0,1/2,0)$ | $(x_1,x_2,0,0,0,0)$ | $T_{1u}(x_1)$ | 0.34811 |
| | | $T_{1u}(x_2)$ | 0.44506 |
| | O displacement | | |
| $R_3^+(1/2,1/2,1/2)$ | $(a,b)$ | $A_{2u}(a)$ | $-0.04325$ |
| | | $A_{2u}(b)$ | 0.34054 |
| $R_4^+(1/2,1/2,1/2)$ | $(r_1,r_2,r_3)$ | $E_u(r_1)$ | 1.36650 |
| | | $E_u(r_2)$ | $-1.35260$ |
| | | $E_u(r_3)$ | 0.07425 |
| $X_5^+(0,1/2,0)$ | $(x_1,x_2,0,0,0,0)$ | $E_u(x_1)$ | 0.43308 |
| | | $E_u(x_2)$ | 0.39501 |
| $M_2^+(1/2,0,1/2)$ | $(0,a,0)$ | $A_{2u}(a)$ | 0.02312 |
| $M_3^+(1/2,0,1/2)$ | $(0,m,0)$ | $E_u(m)$ | 1.08073 |
| | $Pnma$ structure of LaMnO$_3$ | | |
| | La displacement | | |
| $X_5^+(0,1/2,0)$ | $(x,x,0,0,0,0)$ | $T_{1u}(x)$ | 0.57092 |
| | O displacement | | |
| $R_4^+(1/2,1/2,1/2)$ | $(r,-r,0)$ | $E_u(r)$ | 1.25360 |
| $X_5^+(0,1/2,0)$ | $(x,x,0,0,0,0)$ | $E_u(x)$ | 0.16771 |
| $M_2^+(1/2,0,1/2)$ | $(0,a,0)$ | $A_{2u}(a)$ | $-0.34723$ |
| $M_3^+(1/2,0,1/2)$ | $(0,m,0)$ | $E_u(m)$ | 0.95487 |

As discussed by Carpenter and Howard [32], the symmetry of a perovskite lattice which combines an octahedral tilting and orbital ordering depends also on the way how the orbital pattern enclosed into the tilted structure. A combination of the $a^+b^-b^-$ tilting pattern with the $R_3^+(0,b)$ orbital ordering results in two different structures with monoclinic $P2_1/c$ and triclinic $P\bar{1}$ symmetries [32]. The former corresponds to the case when the Jahn-Teller distortion of octahedron occurs in the plane perpendicular to the axis of the in-phase tilting. The latter accounts the case when the octahedra are distorted in a plane which includes the axis of the in-phase tilting. The primary distortions in the perovskite structure of TlMnO$_3$, namely, $a^+b^-b^-$ octahedral tilting and the $R_3^+(0,b)$ orbital ordering, are combined to meet the second case, and their common action reduces the symmetry down to the triclinic $P\bar{1}$. This symmetry allows the tilting pattern to be more general $a^{\pm}b^{-}c^{-}$ through a coupling to secondary displacement modes, whose amplitudes however are very small (Table I and Table 2S of the Supplemental Material [27]).

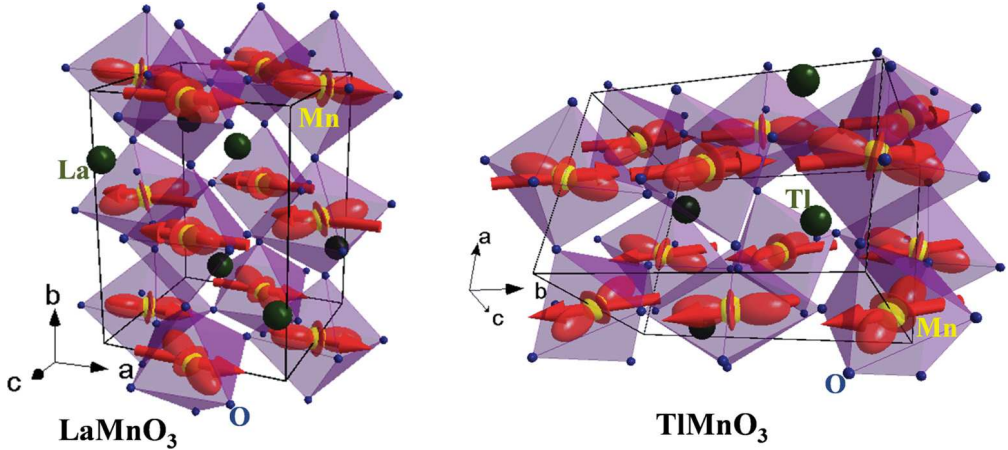

FIG. 2. Schematic representation of the orbital ordering and magnetic structure for LaMnO$_3$, taken from the MAGNDATA database of the Bilbao Crystallography server (#0.1) [26] (left) and for TlMnO$_3$, obtained in this work (right).

### B. Magnetic structure

Below $T_N \sim 92$ K, a set of additional reflections indexed by the $\mathbf{k} = (1/2, 0, 1/2)$ propagation vector (referring to the triclinic cell) appears, indicating onset of a long-range magnetic order [Fig. 3 (upper)]. This propagation vector is consistent with the A-type magnetic structure expected based on the orbital ordering discussed in the previous section. The Goodenogh-Kanamori rules for 180-degree superexchange interactions via oxygen imply a ferromagnetic coupling within the $(1, 0, \bar{1})$ planes and antiferromagnetic between them [28]. The structure has been also found to be the ground state of TlMnO$_3$ in the recent hybrid density functional calculations [20]. The quantitative structure refinement confirmed the A-type spin arrangement and revealed that the spins to a first approximation are confined within the $(1, 0, \bar{1})$ plane, making the angle of $\sim 48°$ with the $b$ axis [Figs. 2 and 4 (right)]. The value of the moments constrained in the final refinement to be collinear and the same on the all Mn sites, was found to be $m = 3.71(2)\mu_B$ ($m_a = 1.50(2)\mu_B, m_b = 2.45(2)\mu_B$, and $m_b = 2.16(2)\mu_B$) at $T = 1.5$ K and its temperature dependence is shown in the inset of Figure 3 (bottom). The model provides a very good refinement quality [Fig. 3 (bottom)] and implies the $P_S\bar{1}$ magnetic symmetry. A description of the crystal and magnetic structures in this space group is given in Table 3S of the Supplemental Material [27]. A decomposition of the magnetic structure of TlMnO$_3$ as well as the magnetic structure of LaMnO$_3$, taken from the MAGNDATA database (#0.1) of the Bilbao crystallographic server [26], in terms of symmetry adapted magnetic modes are given in Table II (only primary magnetic modes are listed, the full list of the symmetry allowed magnetic modes can be found in Table 4S (for TlMnO$_3$) and Table 5S (for LaMnO$_3$) of the Supplemental Material [27]).

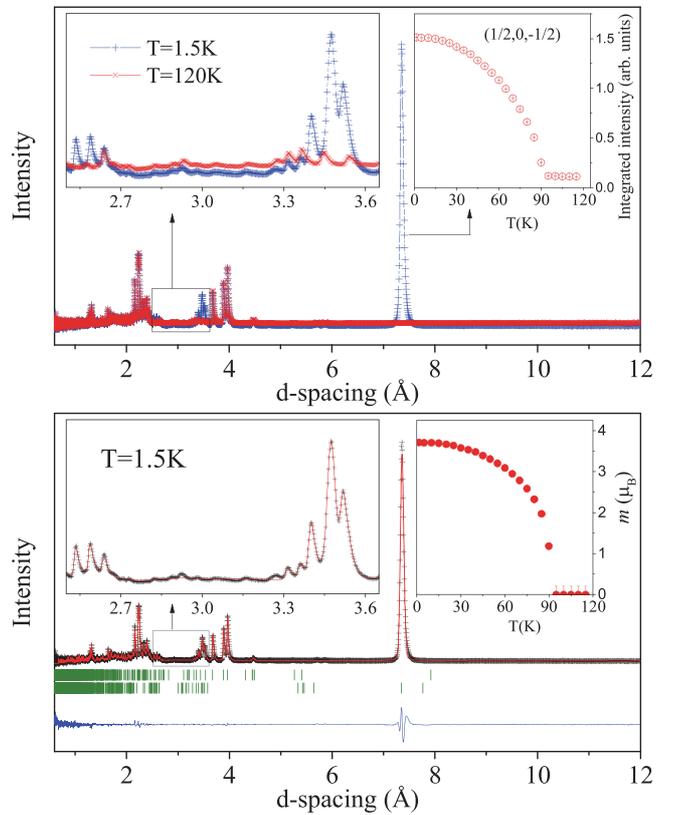

FIG. 3. Neutron diffraction patterns (detector bank with average $2\theta = 58°$) collected at different temperatures (upper). Insets show magnified region, where magnetic reflections are observed (left) and integrated intensity of the $(1/2, 0, -1/2)$ magnetic reflection as a function of temperature (right). Rietveld refinement ($R_{Bragg(nuclear)} = 3.93\%$, $R_{magnetic} = 2.05\%$) of the low temperature neutron diffraction pattern (bottom). The cross symbols (black) and solid line (red) represent the experimental and calculated intensities, respectively, and the line below (blue) is the difference between them. Tick marks indicate the positions of Bragg peaks for the nuclear (top) and magnetic (bottom) scattering. Insets show magnified region with the magnetic reflections (left) and temperature dependence of the magnetic moment (right).

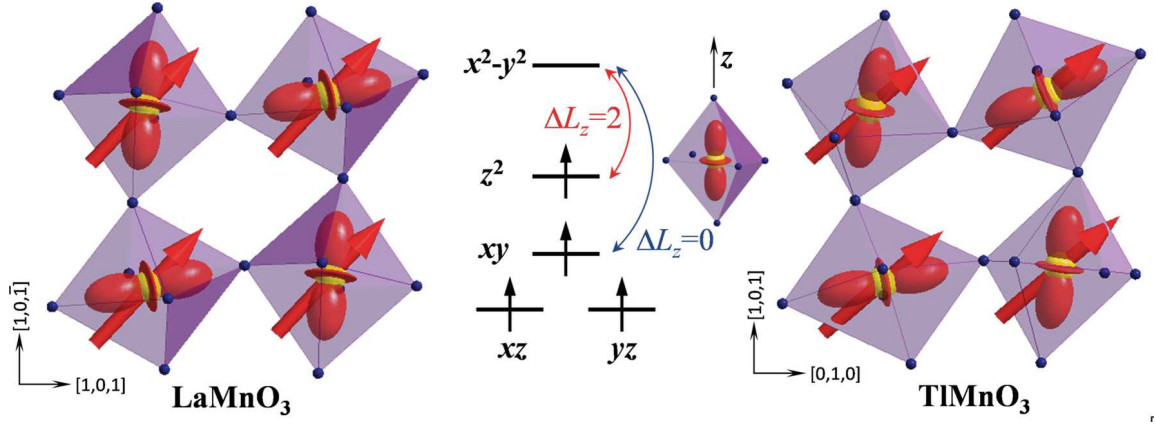

FIG. 4. A fragment of the $(0,1,0)$ plane for LaMnO$_3$ (left) and the $(1,0,\bar{1})$ plane for TlMnO$_3$ (right) containing the Jahn-Teller distortions and spins in the primary A-type spin configuration. In the middle, a schematic diagram illustrating the crystal field splitting of $3d$ electronic levels of Mn$^{3+}$ by an octahedron with four comparable and two essentially longer Mn-O distances along the local $z$ axis [see Fig. 1 (inset) and Fig. 2S of the Supplemental Material [27]].

Comparing the magnetic structures of LaMnO$_3$ and TlMnO$_3$ (Figs. 2 and 4), one can see that in both cases, the orbital patterns define not only the relative coupling between the spins (symmetric exchange) but also their direction in the structure (magnetic anisotropy). The spins are close to the planes containing the Jahn-Teller distortions induced by the alternating $d_{z^2}$ orbitals and bisect the angle made by the orbitals localized on the neighbor Mn sites (Fig. 4). This spin direction can be understood by considering the single ion anisotropy of Mn$^{3+}$ in octahedra elongated along the local $z$ axes due to the Jahn-Teller distortions. An electronic diagram illustrating the crystal field splitting of the $3d$-energy levels in an elongated octahedron is shown in Fig. 4 (middle). For simplicity, we took into account only the dominant distortion of the octahedron. The first excited state which allows a spin-orbit interaction involves an electronic transfer from the $d_{xy}$ to $d_{x^2-y^2}$ orbitals. This electronic transfer does not change the magnetic quantum number ($\Delta L_z = 0$) resulting in the easy axis type of anisotropy, as suggested by the qualitative perturbation theory proposed recently by Whangbo et al. [33]. Thus, the single ion anisotropy favors a noncollinear magnetic structure with the local easy axes of octahedra being along the longest Mn-O bonds (imposed by the Jahn-Teller distortions). The symmetric Heisenberg exchange interactions (superexchange) however require a collinear state. The experimentally observed nearly collinear magnetic structure (see Figs. 2 and 4) minimizes the Hamiltonian which includes the anisotropic nearest neighbor exchange interactions $J_{ij}$ and the single ion anisotropy terms:

$$H = \sum_{ij} J_{ij} \boldsymbol{S}_i \boldsymbol{S}_j - (\boldsymbol{D}_i \boldsymbol{S}_i)^2 - (\boldsymbol{D}_j \boldsymbol{S}_j)^2, \quad (2)$$

where $\boldsymbol{D}_i$ and $\boldsymbol{D}_j$ are vectors along the local easy axes. The spins on the neighbor Mn sites $\boldsymbol{S}_i$ and $\boldsymbol{S}_j$ are expected to be almost collinear in the limit when the exchange interactions are substantially bigger than the energy of the single ion anisotropy $J_{ij} \boldsymbol{S}_i \boldsymbol{S}_j \gg (\boldsymbol{D}_i \boldsymbol{S}_i)^2 + (\boldsymbol{D}_j \boldsymbol{S}_j)^2$.

Let us briefly discuss the antisymmetric Dzyaloshinskii-Moriya (DM) exchange interactions, which are also anisotropic and depend on the direction taken by spins in a crystal structure. We will follow the procedure used in

TABLE II. Decomposition of the magnetic $P_S\bar{1}$ structure of TlMnO$_3$ and $Pn'ma'$ structure of LaMnO$_3$ in respect of the symmetrized magnetic modes of the parent cubic $Pm\bar{3}m$ perovskite structure. The column "Irrep (**k**)" shows the irreducible representations ($m$ indicates time-odd nature of the irreps) of the $Pm\bar{3}m$ space group and the arms of the wave vector star involved. The column "Order parameter" lists the directions of the order parameters in the corresponding irreducible subspaces. The column "Site irrep(var)" displays the point group symmetry irreps of the local Wyckoff positions and one of the free variables from the order parameter directions (in brackets). "Amplitude" shows the amplitudes of the magnetic modes (in $\mu_B$). We provide the standard supercell-normalized amplitudes ($A_S$) defined as the square root of the sum of the squares of the atomic magnetic moments within the primitive supercell.

| Irrep (**k**) | Order parameter | Site irrep(var) | Amplitude |
|---|---|---|---|
| | $P_S\bar{1}$ structure of TlMnO$_3$ | | |
| $mX_3^+(1/2,0,0),(0,0,1/2)$ | $(0,a_2,a_3)$ | $T_{1g}(a_2)$ | 0 |
| | | $T_{1g}(a_3)$ | $-1.43782$ |
| $mX_5^+(1/2,0,0),(0,0,1/2)$ | $(0,0,a_3,a_4,a_5,a_6)$ | $T_{1g}(a_3)$ | 0 |
| | | $T_{1g}(a_4)$ | 0 |
| | | $T_{1g}(a_5)$ | $-7.45737$ |
| | | $T_{1g}(a_6)$ | $-6.70100$ |
| | $Pn'ma'$ structure of LaMnO$_3$ | | |
| $mX_5^+(0,1/2,0)$ | $(a_1,-a_1,0,0,0,0)$ | $T_{1g}(a_1)$ | $-7.49540$ |

Refs. [34–36] to analyze DM interactions in some $Fe^{3+}$-based distorted perovskites. The procedure exploits the fact that antisymmetric exchange is a coupling phenomenon imposed by structural distortions. The analysis is based on trilinear free-energy terms invariant in respect of supergroup symmetry (the cubic $Pm\bar{3}m$ in the present case), where the orthogonal spin modes are transformed by different irreducible representations. These energy terms combine the spin modes with the appropriate structural distortions (atomic displacement modes) which are responsible for the coupling.

As was discussed in the previous section, the dominant structural distortions in $TlMnO_3$ are the rotations of oxygen octahedra which are well approximated by the $a^+b^-b^-$ tilting pattern. The same distortions also dominate in the crystal structure of $LaMnO_3$ (see Table I). In both structures, the out-of-phase octahedral tilting takes place about the $a$ axis and the in-phase tilting about the $b$ axis (in the corresponding triclinic and orthorhombic settings). The axial nature of the distortions implies that the relevant components of the Dzyaloshinskii vector are along the tilting axes and therefore the direction of the spins in the primary magnetic mode, optimizing the antisymmetric exchange, is along the $c$ axis [36]. The experimentally found spin structures in both $TlMnO_3$ and $LaMnO_3$ perovskites indicate that the DM interactions do not control the spin direction, confirming our previous conclusion about the dominant role of the single ion anisotropy. In the case of the $TlMnO_3$ perovskite, however, the spins slightly deviate from the $(1,0,\bar{1})$ plane towards the $c$ axis (about $8°$) pointing to a contribution of the DM interactions to the magnetic anisotropy of the system.

In some cases, the antisymmetric exchange couples a secondary ferromagnetic component resulting in a macroscopic magnetization of antiferromagnetic crystals (weak ferromagnetism). The A-type antiferromagnetic structure is transformed by the six-dimensional $mX_5^+(a_1,a_2,a_3,a_4,a_5,a_6)$ and three-dimensional $mX_3^+(\alpha_1,\alpha_2,\alpha_3)$ time odd irreps of $Pm\bar{3}m$ associated with the $\{k_c = 0,1/2,0\}$ propagation vector star (see Table II). The general free-energy invariants which couple a ferromagnetic component transformed by the $m\Gamma_4^+(f_1,f_2,f_3)$ irrep ($k_c = 0$) with the A-type antiferromagnetic modes are:

$$f_1a_3x_4 - f_1a_4x_3 + f_2a_1x_2 - f_2a_2x_1 + f_3a_5x_6 - f_3a_6x_5 \quad (3)$$

for $mX_5^+$ and

$$f_1\alpha_1x_1 - f_1\alpha_3x_6 - f_2\alpha_2x_4 + f_2\alpha_3x_5 - f_3\alpha_1x_2 + f_3\alpha_2x_3 \quad (4)$$

for $mX_3^+$. These terms require a structural distortions transformed by the time-even $X_5^+(x_1,x_2,x_3,x_4,x_5,x_6)$ irrep to maintain the time-reversal and translational invariance. These distortions with substantial amplitudes exist in the crystal structures of both $TlMnO_3$ and $LnMnO_3$. The order parameter in both structures takes the $X_5^+(x_1,x_2,0,0,0,0)$ direction, but in the latter case, $x_1 = x_2$ (see Table I). The first two nonzero components of the order parameter indicate that the $k_c = (0,1/2,0)$ arm of the propagation vector star is involved in both cases. The order parameters in the primary A-type spin configuration take the directions $mX_5^+(a_1,a_2,0,0,0,0)$ with $a_1 = a_2$ for $LaMnO_3$ and $mX_5^+(0,0,a_3,a_4,a_5,a_6)/mX_3^+(0,\alpha_2,\alpha_3)$ for $TlMnO_3$ (Table II). The latter directions imply that the $k_c = (1/2,0,0)$ and $k_c = (0,0,1/2)$ arms of the propagation vector star are active.

A substitution of these magnetic order parameters into the general expressions (3) and (4) indicates that the $X_5^+$ structural distortions combined with the $mX_5^+/mX_3^+$ A-type primary antiferromagnetic configuration couple a weak ferromagnetic component in the case of $LaMnO_3$ but the invariants vanish in the case of $TlMnO_3$. This result is in a full agreement with the experimentally deduced $\bar{1}1'$ magnetic point group which forbids a macroscopic magnetization. Thus, the structural distortions and the primary magnetic ordering, which potentially can couple a ferromagnetic component in $TlMnO_3$, exploit the different arms of the propagation vector star, and therefore the coupling is forbidden by the translation symmetry.

Note that the $x_i(i=1...6)$ components of the $X_5^+$ order parameter in the coupling terms (3) and (4) can be replaced by the products of $r_i$ and $m_i$ components of the $R_4^+$ and $M_3^+$ order parameters with identical transformational properties as specified by the expression (1). The obtained fourth degree free-energy invariants describe the contribution to the weak ferromagnetism, coming from the common action of the in-phase and out-of-phase tilting. In spite of the fact that these are higher-order terms, they involve the largest structural distortions and therefore can be comparable, in the case of $LaMnO_3$, with the third power terms (3) and (4).

## IV. CONCLUSIONS

The crystal structure of $TlMnO_3$ perovskite possesses triclinic symmetry with the $P\bar{1}$ space group, which is a result of the presence of the $a^+b^-b^-$ octahedral tilting and the antiferrodistortive orbital ordering of the Jahn-Teller active $Mn^{3+}$ cations. Contrary to the family of $LnMnO_3$ manganites, where the Jahn-Teller distortions of octahedra occur in the plane perpendicular to the axis of the in-phase tilting, the structure of $TlMnO_3$ involves the Jahn-Teller distortions in the plane which includes the axis of the in-phase tilting. The primary structural distortions couple a set of secondary displacement modes, resulting in a more general $a^{\pm}b^-c^-$ tilting pattern.

A long range magnetic ordering with the $\boldsymbol{k} = (1/2,0,1/2)$ propagation vector and the $P_S\bar{1}$ symmetry takes place below $T_N \sim 92$ K and involves antiferromagnetic coupling between the ferromagnetic $(1,0,\bar{1})$ planes. This A-type magnetic structure has been theoretically predicted to be stabilized by the anisotropic exchange interactions imposed by the orbital ordering [20]. The spins in the primary magnetic mode are confined close to the $(1,0,\bar{1})$ plane with the value of $3.71(1)\mu_B$ at $T = 1.5$ K. The spin direction is dominated by the single ion anisotropy with the local easy axes defined by the Jahn-Teller distortions of octahedra.

A common action of the in-phase and out-of-phase octahedral tilting as well as the structural distortions trilinearly coupled to them ($X_5^+$ displacement modes) are responsible for the weak ferromagnetic properties in $LnMnO_3$ manganites with the A-type magnetic structure. These displacement modes exist also in the structure of $TlMnO_3$, but in this case, the A-type spin configuration and the structural distortions belong to different arms of the propagation vector star, which implies that the coupling with a macroscopic ferromagnetic component is forbidden by the translation symmetry.

## ACKNOWLEDGMENTS

The work done at ISIS was supported by the project TUMOCS. This project has received funding from the European Union's Horizon 2020 research and innovation programme under the Marie Skłodowska-Curie Grant No. 645660. The work done at NIMS was supported by World Premier International Research Center Initiative (WPI Initiative, MEXT, Japan).